\documentstyle[aps,pre,multicol]{revtex}
\input psfig

\begin{document}

\title{The Free Energy Surface of Supercooled Water}

\author{A. Scala$^{1,2}$, F.~W. Starr$^{1,3}$, E. La Nave$^1$, 
H.~E. Stanley$^1$, and F. Sciortino$^2$}

\address{$^1$Center for Polymer Studies, Center for Computational
Science, and Department of Physics, Boston University, Boston, MA
02215 USA}

\address{$^2$Dipartmento di Fisica e Istituto Nazionale per la Fisica
della Materia, Universit\'{a} di Roma ``La Sapienza'', Piazzale Aldo
Moro 2, I-00185, Roma, Italy}

\address{$^3$Polymers Division and Center for Theoretical and
Computational Materials Science, National Institute of Standards and
Technology, Gaithersburg, MD 20899, USA}

\date{28 July 2000}

\maketitle

\begin{abstract}

We present a detailed analysis of the free energy surface of a well
characterized rigid model for water in supercooled states. We propose a
functional form for the liquid free energy, supported by recent
theoretical predictions [Y. Rosenfeld and P. Tarazona, Mol. Phys. {\bf
95}, 141 (1998)], and use it to locate the position of a liquid-liquid
critical point at $T_{C'} = 130 \pm 5$~K, $P_{C'}=290\pm 30$MPa, and
$\rho_{C'} = 1.10 \pm 0.03$~g/cm$^3$. The observation of the critical
point strengthens the possibility that SPC/E water may undergo a
liquid-liquid phase transition. Finally, we discuss the possibility that
the approach to the liquid-liquid critical point could be pre-empted by
the glass transition.

\end{abstract}

\pacs{PACS numbers: 05.70.Ce, 64.70.Ja, 64.70.Pf }

\begin{multicols}{2}

\section{introduction}

The thermodynamic description of supercooled water has been a major
research topic in recent years. Striking anomalies---such as the
existence of a minimum in the isothermal compressibility $K_T$ along
isobars, the increase of the isobaric specific heat $C_P$ on cooling,
and the temperature of maximum density $T_{MD}$ along
isobars---characterize the behavior of liquid
water~\cite{angell81,debenedetti,ms-rev}. In particular, the study of
supercooled states of water sheds light on the understanding of the
anomalous behavior of liquid water. The increase of $K_T$ and $C_p$ on
supercooling reinforces the possibility that the thermodynamic
properties of supercooled water could be different from those of simple
liquids. Speedy and Angell proposed a scenario in which the increase of
$K_T$ and $C_p$ is related to a re-entrant spinodal line in the phase
diagram of water by postulating the existence of an ultimate limit of
stability for the liquid phase on cooling~\cite{spinodal}.

More recently, increased computing power has made possible the numerical
study of the thermodynamic properties of models for water. In
particular, supercooled states, where relaxation times increase by
several orders of magnitude over typical liquid values, have become
computationally accessible.  It has been shown that explicit atom models
(such as ST2~\cite{st2}, TIP4P/TIP5P~\cite{tips}, and
SPC/E~\cite{spce}), as well as lattice~\cite{lattice} and
continuum~\cite{continuum} models, are able to reproduce the anomalous
thermodynamic properties of water. In all the atomistic models that have
been studied, it has been found that the spinodal line is {\it not}
re-entrant. Additionally, for the ST2 model, the existence of a novel
liquid-liquid critical point has been directly observed in molecular
dynamics simulations~\cite{pses}. Hence, it has been proposed that the
anomalous thermodynamic properties of liquid water could be related to a
liquid-liquid phase transition. According to this hypothesis, two
distinct forms of liquid water, separated by a first-order transition,
may exist below a critical temperature $T_c$; such a critical point
would account for the unusual increases in the thermodynamic response
functions on cooling.  Unfortunately, in water, the estimated $T_c$ is
below the homogeneous nucleation temperature, i.e., inside the so-called
``no-man's land''. This notwithstanding, recent
experiments~\cite{mishima} have probed the possible thermodynamic
scenarios which characterize liquid
water~\cite{ms-rev,spinodal,pses,singfree}.

From a simulation point of view, the ST2 model is the only one that
allows a direct study of the liquid-liquid critical point; an increase
of many orders of magnitude in computing power is needed for a direct
detection of a critical point in other point charge models.  Also, in
supercooled states at the same $T$ and $P$, ST2 molecules are more
mobile compared to real water.  This feature has been exploited for
equilibrating configurations at relatively low
$T$~\cite{pses,paschek-geiger}. The ST2 potential is {\it
over-structured\/} compared to water, and the equation of state is
shifted to higher values of pressure $P$, and temperature
$T$~\cite{pses}.

Among the molecular potentials which have been studied in detail, a
significant role has been played by the extended simple point charge
(SPC/E) model, both because of its simplicity and its success in
capturing the properties of water in the bulk
state~\cite{hpss,sgtc,sss,ChemPhysShort}, as well as in biological
systems~\cite{mounir}. The SPC/E model has three point charges, located
at the atomic centers of the water molecule.  SPC/E is {\it under
structured}, with its equation of states shifted to lower values of P
and T compared to water~\cite{hpss}.  Also, in the supercooled regime,
at the same $T$ and $P$, SPC/E molecules are less mobile than real water
molecules~\cite{sgtc,sss,ChemPhysShort}.  Since it has been shown that
the ST2 and SPC/E models bracket the thermodynamic behavior of water in
the $T-P$ plane~\cite{hpss}, it would be encouraging to clearly detect
the presence of a liquid-liquid critical point also in the SPC/E
potential.  Unfortunately, the reduced diffusivity of SPC/E compared to
ST2 makes it impossible to study directly the low $T$ and high $P$
region, where the SPC/E second critical point should be located.

Here we propose a functional form for the free energy surface of the
SPC/E model in the low temperature region. Our work is supported by
recent theoretical predictions for the $T$ dependence of the potential
energy in supercooled states~\cite{T35}, which have been tested for
several model liquids~\cite{skt,sslss,sri2k,condpoole}.  The calculated
functional form provides a good description of the thermodynamic
quantities in the region where simulations are feasible, and predicts
the presence of a liquid-liquid critical point $C'$ at $T_{C'} = 130 \pm
5$~K, $P_{C'} = 290 \pm 30$MPa, and $\rho_{C'} = 1.10 \pm
0.03$~g/cm$^3$, in reasonable agreement with prior estimates~\cite{hpss}
based on the characteristic shift in thermodynamics properties between
the SPC/E and the ST2 model.

\section{The SPC/E Helmholtz Free Energy}

The numerical data used to calculate the Helmholtz free energy $F = E -
TS$ are obtained from the long molecular dynamics simulations of
Ref.~\cite{sss} for 42 different state-points, comprising 7 different
densities and 6 different temperatures. The simulation results for the
total energy $E$ and pressure $P$ are used here to reconstruct $F$ in
the region $T>210$K, as we describe below.  As noted in
Ref.~\cite{hpss}, the energy $E$ as a function of $V$ develops an
increasingly pronounced convexity on lowering $T$. This signals the
possibility of a phase transition, as $F=E-TS$ will be then also convex
at low $T$.

\begin{figure}[htbp]
\begin{center}
\mbox{\psfig{figure=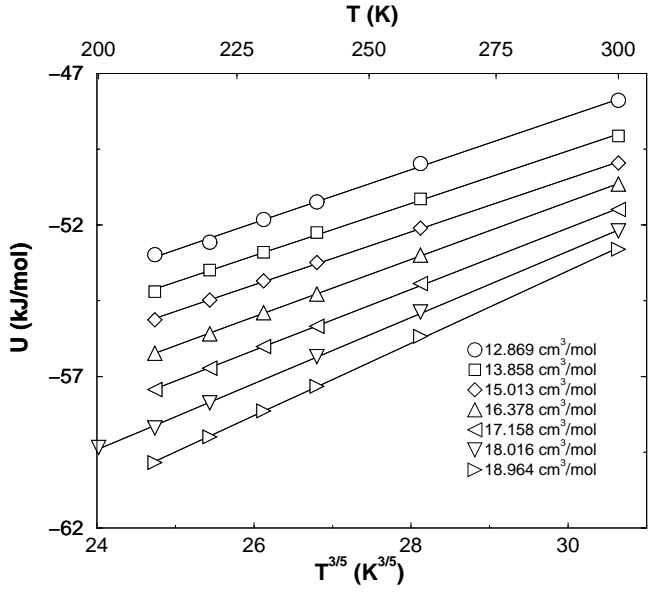,angle=-90,width=3.35in}}
\end{center}
\begin{minipage}{3.35in}
\caption{Fit of the potential energy along isochores with the functional
form $U_0 + \alpha T^{3/5}$.  Symbols denote different molar
volumes. For sake of clarity, the different isochores are shifted by -1
kJ/mol each in order to avoid overlaps.}
\end{minipage}
\label{fig:T35}
\end{figure}

\begin{figure}[htbp]
\begin{center}
\mbox{\psfig{figure=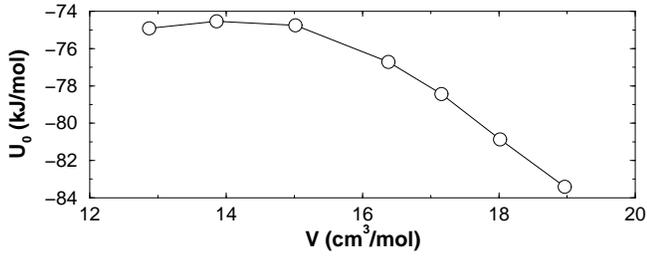,angle=-90,width=3.35in}}
\end{center}
\begin{minipage}{3.35in}
\caption{The volume dependence of $U_0(V)$ from eq.~(\ref{eq:ufit}).
Note that this coincides with $F(T=0$~ $\mbox{K},V)$.  The negative
curvature implies the presence of an unstable region in the phase
diagram at low $T$.}
\end{minipage}
\label{fig:fitV}
\end{figure}

Simulations of the SPC/E model below $T\approx 200$K are not feasible at
the present time, as the time needed to observe equilibrium metastable
properties exceeds currently available resources. Here, the simulation
data for SPC/E water are limited to the region $T>200$K. To investigate
the phase behavior at lower $T$, we exploit the recently-proposed
relationship for the low-$T$ dependence of the potential energy $U$
along isochoric paths~\cite{T35}. Specifically, the low-$T$ behavior of
the potential energy is predicted to follow the functional
form~\cite{T35}
\begin{equation}
U_{\mbox{\scriptsize fit}}(T, V) = U_0(V) + \alpha(V)T^{3/5}.
\label{eq:U(T)}
\end{equation}
Here $U_0$ represents the $T=0$~K value of $U_{\mbox{\scriptsize fit}}$,
which for a classical system may also be identified with $F(0,V)$. The
functional form of Eq.~(\ref{eq:U(T)}) has been shown to describe the
temperature dependence of the potential energy in several different
models, ranging from Lennard-Jones to Yukawa
potentials~\cite{skt,sslss,ssss,sri2k,condpoole}. Although no specific
prediction has been presented until now for molecular systems, we find
that in the temperature range between $T=200$ and $T=300$ K the SPC/E
potential energy is very well described by the $T^{3/5}$ law, as shown
in Fig.~\ref{fig:T35}. The volume dependence of $U_0(V)$ and $\alpha(V)$
are reported in Fig.~\ref{fig:fitV} and in Table~\ref{table:T35}. Since
$F(T=0,V)$ coincides with $U_0(V)$, the clear negative concavity of
$U_0(V)$ at large volumes indicates that if the $T^{3/5}$ law would hold
down to $T=0$K, then the extrapolated liquid free energy would imply a
two-phase coexistence at zero temperature. As will be discussed in more
detail later, at $T=0$K the two phases are separated by a first order
transition around $P=380$ MPa.

Since the $V$ dependence of $U_0$ and $\alpha$ is smooth, we derive a
functional form $U_{\mbox{\scriptsize fit}}(V,T)$, by fitting the values
of $U_0(V)$ and $\alpha(V)$ with sixth degree polynomials
$U_0(V)=\sum_{n=0}^6b_nV^n$ and $\alpha(V)=\sum_{n=0}^6a_nV^n$. We thus
obtain
\begin{equation}
U_{\mbox{\scriptsize fit}}(T,V)=\sum_{n=0}^6 a_n V^n +
 T^{3/5} \sum_{n=0}^6 b_n V^n 
\label{eq:ufit}
\end{equation}

\begin{figure}[htbp]
\begin{center}
\mbox{\psfig{figure=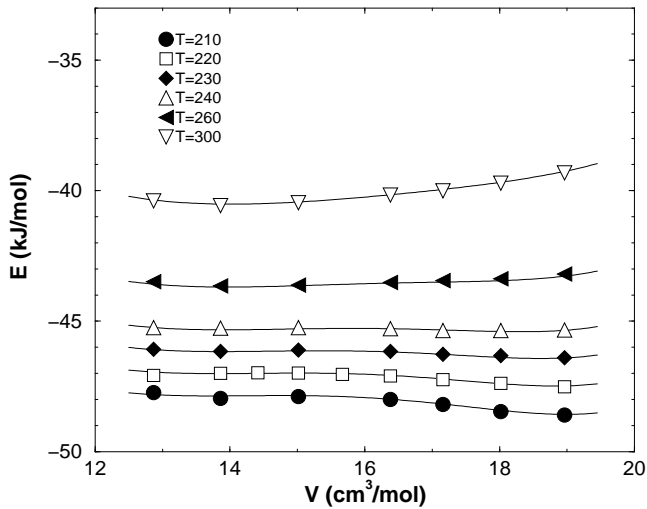,angle=-90,width=3.35in}}
\end{center}
\begin{minipage}{3.35in}
\caption{Comparison between the energy E calculated from simulations
\protect\cite{hpss} and from the free energy approach described here.}
\end{minipage}
\label{fig:Ecompare}
\end{figure}

The $a_n$ and $b_n$ values are reported in Table~\ref{table:polynomial}.
We find almost identical values of $U_0$ and $\alpha$ if we truncate
Eq.~(\ref{eq:ufit}) at order $V^5$. The resulting expression $E(V,T)=3k_B T
+ U_{\mbox{\scriptsize fit}}(V,T)$ for the total energy describes very
well the simulation results, as shown in Fig.~\ref{fig:Ecompare}.

We obtain the entropy $S$ using the thermodynamic relation
\begin{equation}
S(T,V) = S(T_0,V_0) + \frac{1}{T} \int^{T,V}_{T_0,V_0} (dE + P dV) 
\label{eq:dS}
\end{equation}
where the state point $(T_0,V_0)$ is some reference state point.  We
calculate the temperature-dependence of $S$ along isochores from
Eq.~(\ref{eq:ufit}), by performing thermodynamic integration along
constant $V$ paths. $S(T, V)$ is given by
\begin{eqnarray}
\label{eq:S(T)}
S(T, V) &=& S(T_0,V) + \int_{T_0}^{T} \frac{dT}{T} \left(\frac{\partial
E}{\partial T}\right)_V \\
&=& S(T_0,V) + 3 k_B
\ln\left(\frac{T}{T_0}\right) - \frac{3}{2} \alpha(V)
\left(T^{-\frac{2}{5}}-T_0^{-\frac{2}{5}}\right). \nonumber
\end{eqnarray}

The unknown $S(T_0,V)$ function can be evaluated, at any chosen $T_0$,
from the knowledge of the $V$-dependence of $P$ using Eq.~\ref{eq:dS}.
To calculate $S(T_0,V)$ we fit the simulation data for $P(T=300$K$,V)$
again with a sixth-order polynomial
\begin{equation}
P_{\mbox{\scriptsize fit}}(T=300\mbox{K},V) = \sum_{n=0}^6
c_n V^n.
\label{eq:pfit}
\end{equation}
The values of the resulting $c_n$ coefficients are reported in
Table~\ref{table:polynomial}. From elementary calculus,
\begin{eqnarray}
\label{eq:dSy}
S(T_0,V) &=& S(T_0,V_0) -\frac{E(T_0,V_0)-E(T_0,V)}{T_0} + \\
& & \sum_{n=0}^6 \frac{c_n}{n+1}{V^{n+1}-V_0^{n+1}\over T_0}. \nonumber
\end{eqnarray}

The only unknown quantity left is $F(T_0,V_0)$, which can be calculated,
if needed, starting from a known reference point (as for example an
ideal gas state, as done in ref.~\cite{sslss}) and performing
thermodynamic integration up to $V_0,T_0$. The resulting expression for
$F(T,V)= E(T,V)-T S(T,V)$ can then be used to calculate thermodynamic
properties of SPC/E water.

\section{Results}

\begin{figure}[htbp]
\begin{center}
\mbox{\psfig{figure=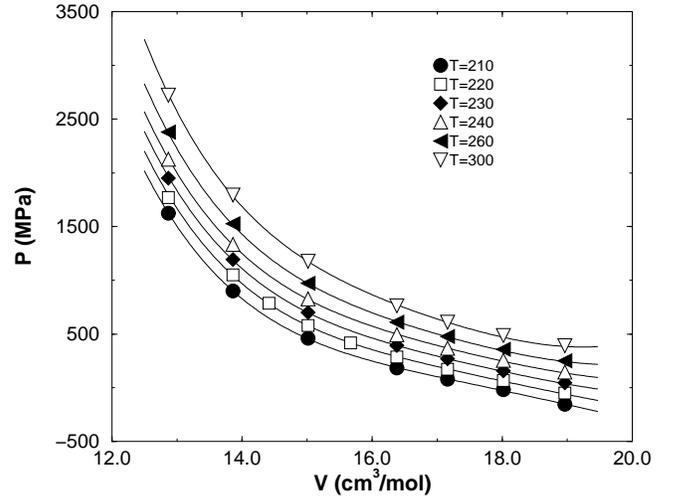,angle=-90,width=3.35in}}
\end{center}
\begin{minipage}{3.35in}
\caption{Comparison between pressures as calculated from
Fig.~\protect\ref{fig:Ecompare}.}
\end{minipage}
\label{fig:Pcompare}
\end{figure}

\begin{figure}[htbp]
\begin{center}
\mbox{\psfig{figure=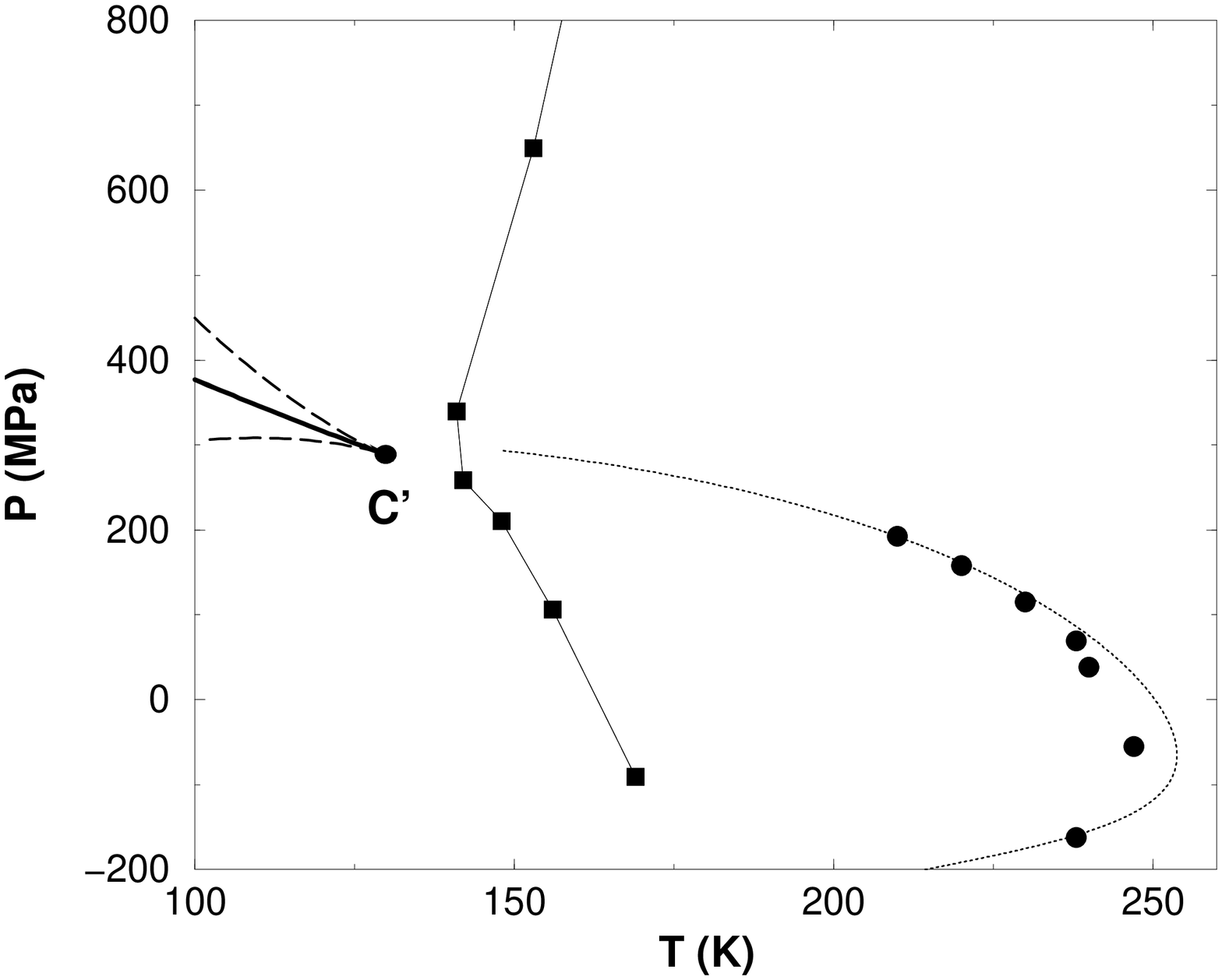,angle=-90,width=3.35in}}
\mbox{\psfig{figure=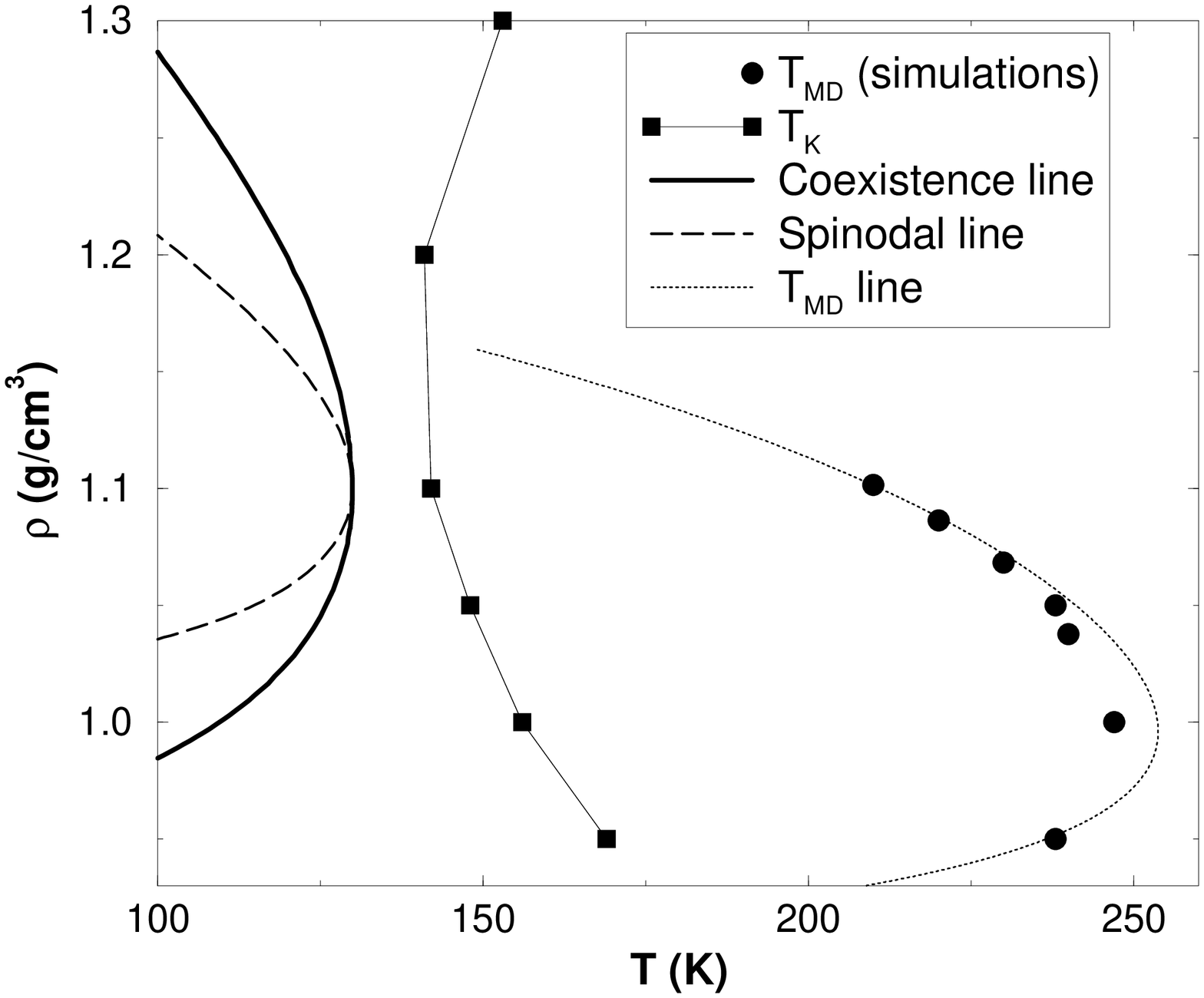,angle=-90,width=3.35in}}
\end{center}
\begin{minipage}{3.35in}
\caption{ P-T (upper panel) and $\rho$-T (lower panel) phase
diagrams. The coexistence line, the spinodals and the $T_{MD}$ line
from our free energy. Squares are $T_{MD}$ points obtained from
simulations; the triangles are the Kauzmann
boundary~\protect\cite{kauzmann} for SPC/E water~\protect\cite{sslss}
below which diffusivity is predicted to
vanish~\protect\cite{adam-gibbsB}.}
\end{minipage}
\label{fig:C-and_TMD}
\end{figure}

First, we compare in Figs.~\ref{fig:Ecompare} and \ref{fig:Pcompare} the
values of $E_{\mbox{\scriptsize fit}}=(1-T\partial/\partial
T)F_{\mbox{\scriptsize fit}}$ and $P_{\mbox{\scriptsize fit}}=-\partial
F_{\mbox{\scriptsize fit}}/\partial V$ with the simulation results for
$T<300$K. Note that the derivatives eliminate the unknown constant
$F(V_0,T_0)$. We also calculate the line of density maxima, $T_{MD}$,
defined as the locus where $(\partial V/\partial T)_P=0$. The predicted
line is compared with the results of the simulations in
Fig.~\ref{fig:C-and_TMD}.

We next use the expression for $F$ to calculate the thermodynamic
properties for $T<200$K where simulations are not feasible.  The free
energy expression proposed depends primarily on the assumption of the
$T^{3/5}$ dependence of the potential energy in supercooled states. The
theoretical prediction and the quality of the $T^{3/5}$ description
reported in Fig.~\ref{fig:T35} suggests that we may meaningfully
extrapolate the calculation to a temperature lower than the one for
which equilibration is feasible at the present time, and search for the
possibility of a liquid-liquid critical point. 

\begin{figure}[htbp]
\begin{center}
\mbox{\psfig{figure=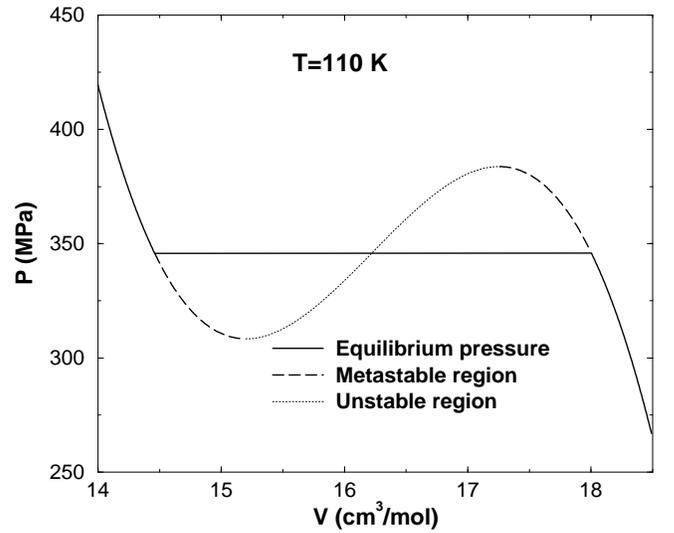,angle=-90,width=3.35in}}
\end{center}
\begin{minipage}{3.35in}
\caption{Pressure at $T=100$K as calculated. The equilibrium pressure is
obtained by the Maxwell construction.}
\end{minipage}
\label{fig:Maxwell-T110}
\end{figure}

We calculate $P_{\mbox{\scriptsize fit}}$ and find that, at temperatures
lower than $130 \pm 5$K, a van der Waals loop
(Fig.~\ref{fig:Maxwell-T110}) develops, signaling a first-order
transition between two liquid phases. The common tangent
construction~\cite{landau} for the Helmholtz free energy
$F_{\mbox{\scriptsize fit}}$ allows us to calculate the coexistence
line; further, we calculate the spinodal lines $(\partial
P_{\mbox{\scriptsize fit}}/\partial V)_T = 0$. The coexistence line
meets the spinodal at a critical point $C'$, which we find at $T_{C'} =
130 \pm 5$K, $P_{C'} = 290 \pm 30$~MPa, and $\rho_{C'} = 1.10 \pm
0.03$~g/cm$^3$.

The resulting SPC/E phase diagram is shown in Fig.~\ref{fig:C-and_TMD}
in both the $(P-T)$ and $(\rho-T)$ planes.  Fig.~\ref{fig:C-and_TMD}
also shows the recently-calculated Kauzmann temperature $T_K$
locus~\cite{sslss}, defined as the temperature at which the
configurational entropy vanishes~\cite{kauzmann}. The evaluation of the
Kauzmann locus is also based on the assumption that the system potential
energy has a $T^{3/5}$-dependence, and hence is fully consistent with
the present free energy calculations. We note that the predicted
critical temperature is $\approx 10$K below the Kauzmann temperature where
SPC/E water is predicted to have a vanishing diffusivity~\cite{sslss}.

As a final consideration, we discuss the interplay between the location
of the critical point and the Kauzmann line. Since at the Kauzmann line
the configurational entropy vanishes, all equilibrium thermodynamic
calculations lose meaning below this line. In this sense, the critical
point in the SPC/E phase diagram should not be considered. In the
potential energy landscape
paradigm~\cite{kauzmann,adam-gibbsB,goldstein,landscape}, the system
would be trapped in a single basin reached at $T_K$. None-the-less, the
free energy below $T_K$ can still be calculated by considering its
separate parts.  The contribution to the free energy due to the
multiplicity of basins sampled would be fixed at its value at $T_K$,
i.e. zero.  Thus, below $T_K$ the intra-basin free energy coincides with
$F$. At low T, frequently, a model based on a harmonic solid is
appropriate for such a calculation~\cite{skt,sslss,ssss}.  The `free energy
calculated will still display a critical point (but slightly shifted
compared to the present estimate) since the basic mechanism which gives
rise to the low-$T$ instability is the shape of $E(V,T)$, which is
already convex well above $T_K$.

\section{Conclusions}

In this article, we have presented a technique of evaluating
thermodynamic quantities in the supercooled region, in a $T$-range where
equilibrium simulations are not feasible due to extremely long
equilibration times. The relevant result of this analysis, applied to
the SPC/E potential, is a clear indication that the free energy surface
develops a region of negative curvature on cooling. A liquid-liquid
critical point develops, in analogy with the behavior of the ST2 model,
for which the location of the critical point is within the region where
equilibrated configurations can be calculated.

The predictions reported in this manuscript are based on a functional
form for the liquid free energy, supported by recent theoretical
predictions~\cite{T35}.  Of course, changes in the temperature
dependence related to novel phenomena which may take place outside the
range where data are available may break the validity of the
extrapolation. In the case of real water, for example, it has been
argued that a change in the $T$-dependence of the thermodynamic
properties takes place in the ``no-man's land''~\cite{angell-nat}.  In
the case of SPC/E water, if such change takes place, it must occur at $T
\lesssim 200 K$, i.e.  in the region where simulations are not
feasible. This would effect our estimate of the location of the critical
point. However, the existence of a region of negative curvature already
in the $T$-region where simulations are feasible supports the
possibility that the liquid-liquid critical transition would take place
at lower temperatures, independently from the assumed $T^{3/5}$ law.

Our results have a particular relevance, since, as previously noticed,
ST2 and SPC/E typically bracket the thermodynamic behavior of the real
liquid. The evidence presented here that the SPC/E potential displays
a critical point at low $T$ and high $P$ strengthens the possibility
that, below the homogeneous nucleation temperature, water may undergo
a liquid-liquid (or glass-glass) phase transition; the two distinct
liquid phases that would appear below $C'$ could correspond to the two
observed amorphous forms of solid water, low density amorphous and
high density amorphous ice.  Indeed, such a transition could be
observed in the glassy state even if $T_{C'} < T_K$, as we find for the
SPC/E model.

The thermodynamic analysis presented here also allows us to grasp the
origin of the presence of the critical point. Indeed, the presence of
the critical point arises from the negative concavity of $E(T,V)$, which
for $T>T_C$ is compensated by the $-TS(T,V)$ contribution. Note that, as
previously observed~\cite{pses}, the negative concavity of $E(T,V)$
already appears in the $T$-region where equilibrium simulations are
feasible, suggesting an inevitable phase transition as the product $TS$
becomes progressively smaller with decreasing $T$. Such negative
concavity of $E(V,T)$ is also found in supercooled
water~\cite{hgk}.

\section{Acknowledgments}
F.W.S. is supported by the National Research Council.  F.S. acknowledges
support from INFM-PRA-HOP, {\it Iniziativa Calcolo Parallelo}, and
MURST-PRIN-98.  This work was also supported by the NSF.

\begin{minipage}{3.35in}
\begin{table}
\caption{Fitting parameters for the potential energy to
Eq.~(\protect\ref{eq:U(T)}). The fits are made for $210$~K~$\le T
\le 300$~K.}
\bigskip
\begin{tabular}{|c|cc|}
$V$ (cm$^3$/mol) & $U_0$ (kJ/mol) & $\alpha$ (kJ/(mol$\cdot$K$^{3/5}$)) \\
\hline
  18.96421  &  -83.41894  &  1.1970260 \\ 
  18.01600  &  -80.86653  &  1.1000960  \\  
  17.15810  &  -78.47431  &  1.0130790  \\ 
  16.37818  &  -76.65199  &  0.9468765 \\  
  15.01333  &  -74.77946  &  0.8767148  \\  
  13.85846  &  -74.50920  &  0.8653371 \\  
  12.86857  &  -74.91184  &  0.8835562 \\  
\end{tabular}
\label{table:T35}
\end{table}
\end{minipage}

\end{multicols}

\begin{table}
\caption{Polynomial fitting coefficients for $U_0(V)$, $\alpha(V)$ (see
Eq.~(\protect\ref{eq:ufit})) and for $P(T=300 K ,V)$
(Eq.~\protect\ref{eq:pfit}). Note that the dimensions of the
coefficients depend on the term $n$ of the expansion.}
\bigskip
\begin{tabular}{|c|ccc|}
$n$ & $a_n$ (kJ$\cdot$mol$^{n-1}$/cm$^{3n}$) & $b_n$ 
(kJ$\cdot$mol$^{n-1}$/(K$^{3/5}\cdot$cm$^{3n}$) & 
$c_n$ (MPa$\cdot$mol$^n$/cm$^{3n}$) \\
\hline
0 & $76.617$               & $-1.8261$              &  $6.8671\times10^{4}$ \\
1 & $-30.435 $             & $0.61927$              &  $2.4466\times10^{3}$\\
2 & $1.279.8 $             & $-2.9301\times10^{-2}$ &  $2.8096\times10^{3}$\\
3 & $5.0719\times 10^{-2}$ & $-9.6397\times10^{-4}$ &  $3.0755\times10^{2}$\\
4 & $4.4964\times 10^{-5}$ & $-2.63884\times10^{-5}$&  $1.2970\times10^{1}$\\
5 & $-3.7530\times 10^{-4}$& $1.07393\times10^{-5}$ &  $0.1871$\\
6 & $1.1301\times 10^{-5}$ & $-3.1252\times10^{-7}$ &  $3.4974\times10^{-4}$
\end{tabular}
\label{table:polynomial}
\end{table}

\end{document}